\documentclass[twocolumn,showpacs,prl,amsmath,amssymb,superscriptaddress]{revtex4-1}
\usepackage{graphicx}% Include figure files
\usepackage{dcolumn}% Align table columns on decimal point
\usepackage{bm}% bold math

\usepackage{xspace,amsmath,amsfonts,amsthm,amssymb,amsbsy,graphics,color,graphicx,bbm}
\usepackage{txfonts}
\usepackage{lmodern}
\usepackage[dvipsnames]{xcolor}
\usepackage{cases}

\usepackage[breaklinks=true]{hyperref}

\hypersetup{
  colorlinks   = true, %Colours links instead of ugly boxes
  urlcolor     = blue, %Colour for external hyperlinks
  linkcolor    = blue, %Colour of internal links
  citecolor   =  red %Colour of citations
}

\newcommand{\bra}[1]{\left\langle{#1}\right|}
\newcommand{\ket}[1]{\left|{#1}\right\rangle}
\newcommand{\op}[2]{\ket{#1}\!\!\bra{#2}}

\newcommand{\ip}[2]{\left\langle{#1}\right|\left.\!{#2}\right\rangle}

\newcommand\tr{\operatorname{tr}}

\def\vec#1{{\bf #1}}

\begin{document}

\title{Fisher-symmetric informationally complete measurements for pure states}

\author{Nan Li}\email{linan@amss.ac.cn}
\affiliation{Center for Quantum Information and Control, University of New
Mexico, Albuquerque NM 87131-0001, USA}
\affiliation{Academy of Mathematics and Systems Science, Chinese Academy of Sciences, Beijing 100190, China}
%\affiliation{Key Laboratory of Random Complex Structures and Data Science, Chinese Academy of Sciences, Beijing 100190, China}
\author{Christopher Ferrie}\email{csferrie@gmail.com}
\affiliation{Center for Quantum Information and Control, University of New
Mexico, Albuquerque NM 87131-0001, USA}
\affiliation{Centre for Engineered Quantum Systems, School of Physics, The University of Sydney, Sydney, NSW, Australia}
\author{Jonathan A.~Gross}\email{jagross@unm.edu}
\affiliation{Center for Quantum Information and Control, University of New
Mexico, Albuquerque NM 87131-0001, USA}
\author{Amir Kalev}\email{amirk@unm.edu}
\affiliation{Center for Quantum Information and Control, University of New
Mexico, Albuquerque NM 87131-0001, USA}
\author{Carlton M.~Caves}\email{ccaves@unm.edu}
\affiliation{Center for Quantum Information and Control, University of New
Mexico, Albuquerque NM 87131-0001, USA}
\affiliation{Centre for Engineered Quantum Systems, School of Mathematics and Physics,
University of Queensland, Brisbane, QLD 4072, Australia}

\date{\today}

\begin{abstract}
We introduce a new kind of quantum measurement that is defined to be symmetric in the sense of uniform Fisher information across a set of parameters that injectively represent pure quantum states in the neighborhood of a fiducial pure state.  The measurement is locally informationally complete---i.e., it uniquely determines these parameters, as opposed to distinguishing two arbitrary quantum states---and it is maximal in the sense of a multi-parameter quantum Cram\'er-Rao bound.  For a $d$-dimensional quantum system, requiring only local informational completeness allows us to reduce the number of outcomes of the measurement from a minimum close to but below $4d-3$, for the usual notion of global pure-state informational completeness, to~$2d-1$.
\end{abstract}

\pacs{03.67.-a, 03.65.Wj}

\maketitle

A fundamental consequence of quantum mechanics is the inability to determine the quantum state of a single physical system.  A well-posed problem in quantum state tomography is this: given many copies of a quantum system, all assumed to be in the same state $\rho$, find a measurement to perform on each copy that is sufficient to specify $\rho$ uniquely in the limit of an infinite number of copies, i.e., from the outcome probabilities.

In quantum theory, measurements are represented by positive-operator-valued measures (\hbox{POVM}s), whose elements, $ E^\xi $, are positive operators satisfying the completeness condition, $\sum_\xi E^\xi = \mathbbm 1$.  If we perform the measurement on a system in state $\rho$, the probability of outcome $\xi$ is $p^\xi=\tr(\rho E^\xi)$.   If the statistics of the outcome probabilities are sufficient to determine all the parameters of the state uniquely, then the \hbox{POVM} is said be a tomographically or  \emph{informationally complete POVM\/} (IC-\hbox{POVM}).

In this paper we consider measurements whose outcome probabilities, though not globally informationally complete, can determine a quantum state in a local neighborhood of a fixed, but arbitrary \emph{fiducial state\/}; we dub such measurements \emph{locally informationally complete}.  We quantify the information content of a measurement using the multi-parameter Fisher-information matrix~\cite{Fi1}.  We look for measurements that satisfy two requirements.  First, the information obtained should be maximal relative to a fundamental bound on the classical Fisher-information matrix, established by Gill and Massar~\cite{GiM99}; by measuring the amount of classical Fisher information relative to the corresponding quantum Fisher information, the Gill-Massar (GM) bound is the multi-parameter expression of the \emph{quantum Cram\'er-Rao bound\/}~\cite{MQ1,Q1,Q2,Q3}.  Saturating the GM~bound requires that the \hbox{POVM} elements be rank one.  Second, we look for measurements such that the classical Fisher information is distributed as uniformly as possible among the parameters of the quantum state.  Measurements that satisfy these requirements are as efficient as possible for measuring all the parameters of a quantum state simultaneously; we call such measurements \emph{Fisher symmetric}.

In this paper we specialize to pure states~\cite{mixed}, where Fisher symmetry means that all the parameters of the pure state are determined with the same resolution relative to the corresponding quantum Cram{\'e}r-Rao bound, and we refer to the measurements that meet our requirements as \emph{pure Fisher-symmetric informationally complete\/} (\hbox{PFSIC}).  We show that $2d-1$ outcomes are necessary and, by example, sufficient for a~\hbox{PFSIC} measurement, in contrast to a minimum close to but below $4d-3$ outcomes required for global pure-state informational completeness~\cite{HMW13}.  Moreover, in accordance with the GM~bound, each of the $2d-2$ parameters of the pure state is determined with a resolution half that of a separate quantum-limited measurement of that \hbox{parameter}.

What is the minimal number of elements for a \hbox{POVM} to be globally informationally complete?  A full-rank quantum state, described by a normalized density operator $\rho$ in a $d$-dimensional Hilbert space, is specified by $d^2-1$ real parameters.  Since the outcome probabilities establish a series of linear constraints for the mixed-state parameters and the \hbox{POVM} operators have to satisfy the completeness condition, it is simple to conclude that an IC-\hbox{POVM} must have at least $d^2$ elements.  If, in addition, one asks for a minimal, rank-one IC-\hbox{POVM} and demands global symmetry in the geometry of the rank-one \hbox{POVM} elements, one arrives at symmetric IC-\hbox{POVM}s (SIC-\hbox{POVM}s)~\cite{Zau99,RBS03,App14}, whose existence in all dimensions is a topic of both mathematical and physical \hbox{interest}.

These considerations change if one knows that $\rho$ is a pure state.  Since the relation between the outcome probabilities and the pure-state parameters is quadratic rather than linear, the problem of the minimal number of elements in a global IC-\hbox{POVM} for pure states is more complicated.  Flammia, Silberfarb, and Caves~\cite{FSC05} considered pure-state informationally \hbox{(PSI-)}complete \hbox{POVM}s, whose outcome probabilities are sufficient to determine a generic pure state (up to a global phase), i.e., all states except for a set of pure states that is dense only on a set of measure zero.  They showed that the minimal number of elements for these \hbox{POVM}s is $2d$, and they conjectured that $2d$ outcomes suffice even if the \hbox{POVM} elements are rank one.  Finkelstein~\cite{Fin04} confirmed this conjecture, but went on to show that if a rank-one \hbox{POVM} achieves global informational completeness for \emph{all\/} pure states, not just a generic set [he called such a \hbox{POVM} pure-state informationally really \hbox{(PSIR-)}complete], then the \hbox{POVM} must have at least $3d-2$ \hbox{POVM} elements.  Finkelstein left open the question of whether a PSIR-complete POVM with this number of outcomes exists.  Recently, in a {\it tour de force\/} of mathematical physics, Heinosaari, Mazzarella, and Wolf~\cite{HMW13} showed that a \hbox{POVM} that identifies all pure states has minimally $4d-3-c(d)\alphaup(d)$ outcomes, where $1\le c(d)\le2$ and $\alphaup(d)$ is the number of 1s in the binary expansion of $d-1$.  This brief discussion illustrates the complicated nature of global informational completeness for pure states.

Here we consider a different tomographic problem, which might be called local or point tomography.  In this problem, an experimenter knows which pure state she is trying to prepare and knows she can prepare it quite well, except that the unitary operators used in the preparation have systematic errors; though the prepared state is pure, it is different from the fiducial state the experimenter is aiming for.  The experimenter wants to find a measurement sufficient to identify all small discrepancies from this fiducial state.  We quantify the ``goodness'' of a measurement with the multi-parameter Fisher-information matrix~\cite{Fi1}.  Fisher information is a key tool in statistics, which allows one to bound the performance of any estimator, and it has played a prominent role in the development of quantum information and metrology, where it has been suitably generalized to the quantum setting~\cite{MQ1,Q1,Q2,Q3}.  It is the appropriate tool here because it provides a definitive answer to questions about estimating local deviations from a fiducial state.

Now we formulate the problem with precision, first generally for mixed states and then for the pure-state context analyzed in this paper~\cite{mixed}.  An unknown quantum state $\rho(\vec x)$ depends on a vector $\vec x=(x^1,\ldots, x^p)$ of $p$ real parameters.  For density operators of rank $\ell$ in $d\ge\ell$ dimensions, $p=2d\ell-\ell^2-1$; for full-rank density operators, $p=d^2-1$, and for pure states, $p=2d-2$.  The fiducial state can be labeled by $\vec x=\vec 0$.  In the following, we are interested in the quantum and classical Fisher-information matrices evaluated at the fiducial state, i.e., at $\vec x=\vec 0$.

The quantum Fisher information is a $p\times p$ real, symmetric matrix $\vec Q(\rho)$, whose matrix elements are
\begin{equation}\label{eq:QFisher}
Q _{\alpha\beta}(\rho)=\frac 12 \tr[\rho(L_{\alpha} L_{\beta}+ L_{\beta} L_{\alpha})],
\end{equation}
with the (Hermitian) \emph{symmetric logarithmic derivative\/} (SLD) operators $L_{\alpha}$, one
for each parameter, determined implicitly by $\partial\rho/\partial x^\alpha =\frac{1}{2}(L_{\alpha}\rho +\rho L_{\alpha})$.  We can reparameterize the quantum state to make $\vec Q$ the identity matrix.

Given a \hbox{POVM}, with elements $E^\xi$ and outcome probabilities $p^\xi=\tr(\rho E^\xi)$, the classical Fisher-information matrix is a $p\times p$ real, symmetric matrix $\bf C$, defined by
\begin{align}\label{eq:CFisher}
C_{\alpha\beta}
=\sum_\xi \frac{1}{p^\xi}\frac{\partial {p^\xi}}{\partial {x^{\alpha}}}
\frac{\partial {p^\xi}}{\partial{x^{\beta}}}.
\end{align}
The sum over $\xi$ here is restricted to $E^\xi$ that are not orthogonal to the fiducial state $\rho$, i.e., for which $p^\xi=\tr(\rho E^\xi)\ne0$.

Gill and Massar~\cite{GiM99} proved that for any states $\rho(\vec x)$ and any \hbox{POVM}, the classical Fisher-information matrix $\bf C$ satisfies
\begin{equation}\label{eq:GMbound}
\tr\!\big(\vec Q^{-1}\vec C\big)=\tr\!\Big(\vec Q^{-1/2}\vec C\vec Q^{-1/2}\Big)\leq d-1,
\end{equation}
with equality if and only if all the \hbox{POVM} elements are rank one, and none is orthogonal to the fiducial state.  The GM~quantity, $\tr(\vec Q^{-1}\vec C)$, is invariant under reparameterization of the quantum states.  It is most easily interpreted when the parameters are chosen so that $\vec Q$ is the identity matrix.  The unit elements on the diagonal of $\vec Q$ then express the quantum limit, called the \emph{quantum Cram\'er-Rao bound}, on estimating each of the parameters separately~\cite{Q3}.  The corresponding diagonal elements of the classical Fisher matrix give the performance of the \hbox{POVM} in determining these same parameters in units of the quantum limit.  The GM~quantity is the sum of the diagonal elements of $\vec C$; the bound~(\ref{eq:GMbound}) expresses the quantum limit on estimating all the parameters simultaneously and as such is a Fisher-information expression of the uncertainty \hbox{principle}.  Zhu~\cite{Zhu2015a} has made use of the positive symmetric matrix $\vec Q^{-1/2}\vec C\vec Q^{-1/2}$ in a study of information complementarity and incompatible observables.

We call a \hbox{POVM} \emph{Fisher symmetric\/} if it saturates the GM bound~(\ref{eq:GMbound}) and has classical Fisher matrix distributed as uniformly as possible among the parameters of the quantum state.  What we mean by as uniformly as possible is that the measurement minimizes the quadratic quantity
\begin{align}\label{eq:FSquantity}
\begin{split}
\tr\!\big(\vec Q^{-1}\vec C\vec Q^{-1}\vec C\big)
&=\tr\!\Big(\big(\vec Q^{-1/2}\vec C\vec Q^{-1/2}\big)^2\Big)\\
&\ge\frac{\Big[\tr\!\Big(\vec Q^{-1/2}\vec C\vec Q^{-1/2}\Big)\Big]^2}{p}
=\frac{(d-1)^2}{p}\;.
\end{split}
\end{align}
The inequality follows directly from minimizing this quadratic quantity subject to the linear trace constraint.   The absolute minimum is achieved if and only if the classical Fisher matrix is proportional to the quantum Fisher matrix, i.e., $\vec C=(d-1)\vec Q/p$.  For full-rank density operators, however, there is generally no POVM that achieves the absolute minimum; it can be achieved only for qubits and for the maximally mixed state in all dimensions~\cite{mixed}.  For any density operator, however, the classical Fisher matrices are a convex set under coin-flipping convex combinations of the underlying POVMs; since the quantity~(\ref{eq:FSquantity}) is convex, global minima are guaranteed to exist.  Study of Fisher symmetry for full-rank density operators will thus be focused on finding what the minimum value is and what POVMs achieve it; such POVMs determine a quantum state locally as efficiently as possible.

For pure states, it is possible to achieve $\vec C=\frac12\vec Q$, as we now show.  A POVM that achieves $\vec C=\frac12\vec Q$, what we call a PFSIC, can estimate all the parameters of a pure state with half the quantum-limited resolution with which each could be estimated separately.  The unknown pure state $\rho(\vec x)=\op{\Psi(\vec x)}{\Psi(\vec x)}$ depends on a vector $\vec x$ of $2d-2$ real parameters.  We denote the fiducial state $\vec x=\vec 0$ by $\ket 0$, i.e., $\ket{\Psi(\vec 0)}=\ket 0$.   By assumption, the unknown pure state $|\Psi(\vec x)\rangle$ is close to $\ket 0$ and thus can be parameterized to linear order as
\begin{equation}
|\Psi(\vec x)\rangle = \ket 0+\sum_{k=1}^{d-1}(x^{k0}+ix^{k1})\ket k.
\end{equation}
where the set $\{\ket k\}_{k=0,1,\ldots,d-1}$ is an orthonormal basis, $\vec x=\big(x^{1,0},x^{1,1},...,x^{d-1,0},x^{d-1,1}\big)$, and $|x^{k\sigma}|\ll 1$ for $k=1,...,d-1$, $\sigma=0,1$.  Keeping only the terms linear in the parameters, we get
\begin{equation}
\rho(\vec x)
=\op 0 0 + \sum_{k,\sigma} x^{k\sigma}X_{k\sigma}
=\op 0 0 + \sum_\alpha x^\alpha X_\alpha,
\label{eq:rhox}
\end{equation}
where
\begin{align}
X_{k\sigma}=(-i)^\sigma\big(\op 0 k + (-1)^\sigma\op k 0\big),
\end{align}
i.e., $X_{k0}=\op 0 k + \op k 0$ and $X_{k1}=-i\op 0 k +i \op k 0$, for $k=1,\ldots,d-1$.  In Eq.~(\ref{eq:rhox}), we ignore second-order terms because they do not contribute either to the quantum Fisher information or to the measurement-induced classical Fisher information at the fiducial point.  In accordance with the foregoing, we sometimes let a single Greek index stand for both $k$ and $\sigma$, as in the last form of  Eq.~(\ref{eq:rhox}).  The Hermitian operators $X_\alpha$ satisfy $\tr(X_\alpha X_\beta)=2\delta_{\alpha\beta}$.

When $\rho(\vec x)$ is a pure state, the SLDs are easy to find.  At the fiducial state, the SLDs are $L_\alpha = 2\partial\rho(\vec x)/\partial x^{\alpha}|_{\vec x=0}=2X_\alpha$, and the quantum Fisher-information matrix is $\vec Q = 4\vec I_{2d-2}$, where $\vec I_n$ denotes the $n\times n$ identity matrix.  We have thus chosen from the start a parameterization that is only a uniform rescaling away from making the quantum Fisher matrix the identity.

Now consider any \hbox{POVM} that saturates the GM~bound, i.e., has $n$ rank-one \hbox{POVM} elements, none of which is orthogonal to $\rho(\vec 0)=\op 0 0$, and take the \hbox{POVM} elements to be
\begin{align}
E^\xi = \op{\psi^\xi}{\psi^\xi}=\sum_{k,j=0}^{d-1}a^{\xi}_k(a^{\xi}_j)^*\op k j,
\end{align}
where the \hbox{POVM} vectors are
\begin{align}\label{eq:POVMvectors}
\ket{\psi^\xi} = \sum_{k=0}^{d-1} a^{\xi}_k \ket k,
\quad\xi=0,\ldots,n-1.
\end{align}
The \hbox{POVM} completeness condition, $\sum_\xi E^\xi = \mathbbm 1$, becomes $\sum_\xi(a^{\xi}_j)^*a^{\xi}_k =\delta_{jk}$.   Defining
\begin{equation}
a^{\xi}_k = b_k^{\xi} + i c_k^{\xi},\qquad k=0,...,d-1,
\end{equation}
and gathering up the various components into $n$-dimensional column vectors,
\begin{align}\label{eq:bc}
\vec{b}_k
=
\begin{pmatrix}
b_k^0\\[2pt]
b_k^1\\[2pt]
\vdots\\[2pt]
b_k^{n-1}
\end{pmatrix},
\quad
\vec{c}_k
=
\begin{pmatrix}
c_k^0\\[2pt]
c_k^1\\[2pt]
\vdots\\[2pt]
c_k^{n-1}
\end{pmatrix},
\quad
k=0,1,\ldots,d-1,
\end{align}
we can put the completeness condition in the form,
\begin{align}\label{eq:completeness}
\begin{split}
\vec b_j\cdot\vec b_k+\vec c_j\cdot\vec c_k &= \delta_{jk},\\
\vec b_j\cdot\vec c_k-\vec c_j\cdot\vec b_k &= 0,
\end{split}
\qquad
j,k=0,1,...,d-1.
\end{align}
Choice of phase of the \hbox{POVM} vectors~(\ref{eq:POVMvectors}) allows us to make $a^{\xi}_0=b^\xi_0$ real and nonnegative (thus $c^\xi_0=0$) for all $\xi$.  Since we now have $\vec c_0=0$, there are $2d-1$ nonzero vectors~(\ref{eq:bc}).  It is useful to spell out separately the $j=0$ or $k=0$ parts of the completeness conditions:
\begin{align}\label{eq:completeness0}
\begin{split}
\vec b_0\cdot\vec b_0&=1,\\
\vec b_0\cdot\vec b_k=\vec b_0\cdot\vec c_k&=0,\quad k=1,\ldots,d-1.
\end{split}
\end{align}

The probability to obtain the outcome $\xi$ at the fiducial state is
\begin{equation}\label{eq:pxi}
p^\xi  = \bra 0 E^\xi\ket 0 = |a^{\xi}_0|^2=(b_0^\xi)^2>0.
\end{equation}
Notice that the vector $\vec b_0$ is a normalized vector that has strictly positive components.
We also have $\partial p^\xi(\vec x)/\partial x^\alpha=\tr(E^\xi X_\alpha)$, i.e.,
\begin{subequations}\label{eq:pxipartials}
\begin{align}
\frac {\partial {p^\xi}}{\partial {x^{k0}}}
&=\bra k E^\xi \ket 0+\bra 0 E^\xi \ket k
=2b^{\xi}_0 b^{\xi}_k,\\
\frac {\partial {p^\xi}}{\partial {x^{k1}}}
&=-i\bra k E^\xi \ket 0+i\bra 0 E^\xi \ket k
=2b^{\xi}_0 c^{\xi}_k.
\end{align}
\end{subequations}

The definition~(\ref{eq:CFisher}) of the classical Fisher matrix gives, for $j,k=1,...,d-1$,
\begin{subequations}\label{eq:Fishermatrix}
\begin{align}
C_{j0,k0}& = 4 \sum_\xi b_j^{\xi}b_k^{\xi}=4\vec b_j\cdot\vec b_k,\\
C_{j1,k1}& = 4 \sum_\xi  c_j^{\xi}c_k^{\xi}=4\vec c_j\cdot\vec c_k,\\
C_{j0,k1}=C_{k1,j0}&=4 \sum_\xi b_j^{\xi}c_k^{\xi}=4\vec b_j\cdot\vec c_k.
\end{align}
\end{subequations}
The classical Fisher matrix is a matrix of inner products of the $2d-2$ $n$-dimensional vectors $\{2\vec{b}_k,2\vec{c}_k\}_{k=1}^{d-1}$.  The rank of such a matrix, called a Gram matrix, is the span of the vectors going into the inner products, so the rank of the classical Fisher matrix is bounded above by $\min(n,2d-2)$.

In the pure-state case, the Fisher-symmetry condition becomes $\vec C=\frac12 \vec Q=2\vec I_{2d-2}$.  To satisfy this condition, ${\bf C}$ must be full rank, i.e., have rank $2d-2$, which implies that $n\ge2d-2$. Since ${\bf C}$ is full rank, it is invertible, and the measurement is locally informationally complete in that it uniquely determines all parameters in the limit of infinitely many measurements.

In the parameterization we are using, the Fisher-symmetric \hbox{POVM} must satisfy
\begin{align}\label{eq:FS}
\begin{split}
\textstyle{\frac12}C_{j0,k0}& = 2\vec b_j\cdot\vec b_k=\delta_{jk},\\
\textstyle{\frac12}C_{j1,k1}& = 2\vec c_j\cdot\vec c_k=\delta_{jk},\\
\textstyle{\frac12}C_{j0,k1} &= 2\vec b_j\cdot\vec c_k=0,
\end{split}
\qquad
j,k=1,...,d-1.
\end{align}
Combining these Fisher-symmetry conditions with the completeness conditions~(\ref{eq:completeness0}), we see that the vectors $\{\sqrt2\vec{b}_k,\sqrt2\vec{c}_k\}_{k=1}^{d-1}$ are a set of $2d-2$ orthonormal vectors in an $n$-dimensional subspace, which is orthogonal to $\vec b_0$.  We can now conclude that $n\ge 2d-1$.

It is clear that \hbox{PFSIC}s exist for all $n\ge2d-1$, since they can be constructed by choosing an $n$-dimensional vector $\vec b_0$ with all positive components and then finding $2d-2$ orthonormal vectors in the subspace of dimension $n-1\ge2d-2$ orthogonal to $\vec b_0$.  In the Supplemental Material~\cite{supp}, we construct a minimal ($n=2d-1$) \hbox{PFSIC} \hbox{POVM} by choosing $\vec b_0$ symmetrically, i.e., $\vec b_0^T=\big(1,1,\ldots,1)/\sqrt{2d-1}$ and using a manifestly symmetric method to construct the remaining vectors.  For qubits this \hbox{PFSIC} reduces to the trine measurement, with the \hbox{POVM} vectors corresponding to outcomes in the equatorial plane of the Bloch sphere.  The trine measurement is locally informationally complete for pure states near the north pole of the Bloch sphere, but it is not globally informationally complete because it cannot distinguish states whose Bloch vectors differ by a sign flip of the $z$ component of the Bloch vector.  This exemplifies the sort of ambiguity that prevents Fisher-symmetric measurements from being globally informationally complete.

In the Supplemental Material~\cite{supp}, we also construct a POVM that comes from flipping a coin with probabilities $p_\chi$ and $p_\tau$ to choose between measuring in one of two orthonormal bases,
\begin{align}\label{eq:chitaubases}
\begin{split}
\ket{\chi^\xi}&=u_0^\xi\ket 0+\sum_{j=1}^{d-1}u_j^\xi\ket j\,,\\
\ket{\tau^\xi}&=-iu_0^\xi\ket 0+\sum_{j=1}^{d-1}u_j^\xi\ket j\,,
\end{split}
\quad
\xi=0,1,\ldots,d-1,
\end{align}
with $u^\xi_0>0$.  With the POVM vectors chosen to be $\ket{\psi^\xi}=\sqrt{p_\chi}\ket{\chi^\xi}$ and $\ket{\psi^{d+\xi}}=\sqrt{p_\tau}\ket{\tau^\xi}$, the Fisher-information matrix~(\ref{eq:Fishermatrix}) is diagonal, with diagonal components $C_{j0,j0}=4p_\chi$ for the estimates of the real parts, $x^{k0}$, of the amplitudes, and $C_{j1,k1}=4p_\tau$ for estimates of the imaginary parts, $x^{k1}$.  When $p_\chi=p_\tau=\frac12$, we have a \hbox{PFSIC}.  Other weightings of the coin give different tradeoffs, within the GM~bound, between determining the real and imaginary parts of the amplitudes.  This example illustrates the sense in which the GM~bound can be thought of as an uncertainty principle for measuring simultaneously the parameters that specify a pure state.  For a qubit, the two bases~(\ref{eq:chitaubases}) correspond to measurements of the Pauli operators $\sigma_x$ and $\sigma_y$.

It is useful to clarify what freedom we have in choosing a \hbox{PFSIC}.  Starting with a minimal \hbox{PFSIC}, i.e., with an orthonormal set $\{\vec b_0,\sqrt 2\vec{b}_k,\sqrt 2\vec{c}_k\}_{k=1,\ldots,d-1}$, of ($2d-1$)-dimensional real vectors, where $\vec b_0$ has all positive components, we can do any (active) orthogonal transformation $O$ to get a new set $\vec B_0=O\vec b_0$ and $\vec B_j=O\vec b_j$, $\vec C_j=O\vec c_j$, for $j=1,\ldots,d-1$.  To get a nonminimal \hbox{PFSIC}, we add additional dimensions to the real vector space and allow $O$ to map into these extra dimensions.  In terms of components, we have
\begin{align}
B^{\xi}_j&=\sum_{\eta=0}^{2d-2} O^\xi_\eta b_j^\eta,\quad j=0,1,\ldots,d-1,\\
C^{\xi}_j&=\sum_{\eta=1}^{2d-2} O^\xi_\eta c_j^\eta,\quad j=1,\ldots,d-1.
\end{align}
Letting $\ket{\phi^\xi}$ be the \hbox{POVM} vectors for the primed real vectors, we have
\begin{align}
\ket{\phi^\xi}=\sum_{j=0}^{d-1}(B_j^\xi+iC_j^\xi)\ket j
=\sum_{\eta=0}^{2d-2} O^\xi_\eta\ket{\psi^\eta}.
\end{align}
Thus our freedom is to do any orthogonal mixing of the \hbox{POVM} vectors, subject to the requirement that $B_0^\xi=\ip{0}{\phi^\xi}=\sum_{\eta=0}^{2d-2}O^\xi_\eta\ip{0}{\psi^\eta}\ne0$ (negative components of $\vec B_0$ can be handled by additional reflections in the real vector space or by rephasing the \hbox{POVM} vectors $\ket{\phi^\xi}$).

How does this compare with the usual freedom for rank-one \hbox{POVM}s?  In complete analogy with the Hughston-Josza-Wootters freedom for pure-state ensemble decompositions of a density operator~\cite{HJW}, for \hbox{POVM}s we are asking for the freedom in writing the unit operator as a sum of rank-one operators.  Generally, that freedom is the ability to mix the \hbox{POVM} vectors with any unitary matrix, which always yields another \hbox{POVM}.  The restriction here is that we can only use real unitaries, i.e., orthogonal matrices, that leave $\vec b_0$ with all nonzero components.

The most complete information we can have of a physical system is its quantum state.  We access this information by making repeated measurements on systems prepared in the same state.  There are two fundamental and practical questions about such measurements: (i)~Which schemes are sufficient to specify the state uniquely? (ii)~Which schemes provide the most information per measurement?  Typically such questions are studied separately, the former using tools of linear algebra and convex geometry and the latter using quantum generalizations of Fisher statistics or other methods.  Here we bring these two lines of questioning together to construct measurements that are minimal, symmetric, and informationally complete, but in a local, Fisher-statistical sense rather than in the global, geometric sense.  We have explicitly constructed measurements with $2d-1$ outcomes that are sufficient to estimate simultaneously all the parameters of pure quantum states near a fiducial state.  Moreover, these measurements provide equal and optimal information about all the parameters.  The quantum price for the simultaneous estimation, in accordance with the GM uncertainty principle~(\ref{eq:GMbound}), is that each parameter is determined with half the sensitivity with which it could be determined \hbox{separately}.

This work was supported in part by US~National Science Foundation Grant Nos.~PHY-1212445 and PHY-1521016 and by US Office of Naval Research Grant No.~N00014-15-1-2167.  NL was also supported by the National Natural Science Foundation of China (Grant Nos.~11405262, 61134008, and 11375259), the National Center for Mathematics and Interdisciplinary Sciences, Chinese Academy of Sciences (Grant No.~Y029152K51), and the Key Laboratory of Random Complex Structures and Data Science, Chinese Academy of Sciences (Grant No.~2008DP173182).  CF was also supported by the Canadian Government through the NSERC PDF program, the IARPA MQCO program, the Australian Research Council via EQuS Project No.~CE11001013, and U.S.~Army Research Office Grant Nos.~W911NF-14-1-0098 and~W911NF-14-1-0103.

\setcounter{equation}{0}

\appendix*

\begin{widetext}
\section{Supplemental Material}

\subsection{I.~Minimal PFSIC}

Here we construct a minimal ($n=2d-1$) PFSIC~\hbox{POVM}.  To avoid bulky row and column vectors, we introduce the natural basis vectors $\{\vec e_\xi\}_{\xi=0,1,\ldots,2d-2}$, where $\vec e_\xi$ has a 1 in the $\xi$th position and zeroes elsewhere.

Our objective is to find $2d-1$ orthonormal vectors, $\{\vec u_\xi\}_{\xi=0,\ldots,2d-2}=\{\vec b_0,\sqrt2\vec b_j,\sqrt 2\vec c_j\}_{j=1,\ldots,d-1}$ in $n=2d-1$ dimensions, with $\vec b_0$ having all positive components.  To take full advantage of the symmetry of a PFSIC, we choose
\begin{align}
\vec b_0=\frac{1}{\sqrt n}\sum_{\xi=0}^{2d-2}\vec e_\xi
\qquad\Longleftrightarrow\qquad
b_0^\xi=\frac{1}{\sqrt n}\,,\quad \xi=0,1,\ldots,2d-2.
\end{align}

Our first step is to project the natural basis vectors orthogonal to $\vec b_0$:
\begin{align}
\vec v_\xi=\vec e_\xi-\vec b_0(\vec b_0\cdot\vec e_\xi)=\vec e_\xi-\frac{1}{\sqrt n}\vec b_0\,,
\qquad
\xi=0,1,\ldots,2d-2.
\end{align}
The resulting vectors $\vec v_\xi$ are orthogonal to $\vec b_0$, but there is one too many for them to be an orthogonal set.  They are, however, symmetrically distributed according to
\begin{align}
\vec v_\xi\cdot\vec v_\eta=\delta_{\xi\eta}-\frac{1}{n}\,.
\end{align}
This means that we can choose one of these vectors, say $\vec v_0$, subtract an appropriate multiple of it from all the other vectors, and obtain an orthonormal set of vectors orthogonal to $\vec b_0$:
\begin{align}
\vec u_\xi=\vec v_\xi-\frac{1}{\sqrt n+1}\vec v_0
=\vec e_\xi-\frac{1}{\sqrt n+1}(\vec e_0+\vec b_0)\,,
\qquad
\xi=1,\ldots,2d-2.
\end{align}

We now choose, for $j=1,\ldots,d-1$,
\begin{subequations}
\begin{align}
\vec b_j=\frac{1}{\sqrt2}\vec u_{2j-1}
=\frac{1}{\sqrt2}\!\left(\vec e_{2j-1}-\frac{1}{\sqrt n+1}(\vec e_0+\vec b_0)\right)
\qquad&\Longleftrightarrow\qquad
b^\xi_j
=\frac{1}{\sqrt2}
\left[\delta^\xi_{2j-1}
-\frac{1}{\sqrt n+1}\!\left(\delta^\xi_0+\frac{1}{\sqrt n}\right)\right]\,,\\
\vec c_j=\frac{1}{\sqrt2}\vec u_{2j}
=\frac{1}{\sqrt2}\!\left(\vec e_{2j}-\frac{1}{\sqrt n+1}(\vec e_0+\vec b_0)\right)
\qquad&\Longleftrightarrow\qquad
c^\xi_j
=\frac{1}{\sqrt2}
\left[\delta^\xi_{2j}
-\frac{1}{\sqrt n+1}\!\left(\delta^\xi_0+\frac{1}{\sqrt n}\right)\right]\,.
\end{align}
\end{subequations}
Plugging these components into the expression for the POVM vectors,
\begin{align}\label{eq:POVMvectors1}
\ket{\psi^\xi}=\sum_{j=0}^{d-1} a^\xi_j\ket j
=\frac{1}{\sqrt n}\ket 0+\sum_{j=1}^{d-1}(b^\xi_j+ic^\xi_j)\ket j\,,
\end{align}
we get
\begin{subequations}
\begin{align}
\ket{\psi^0}&=\frac{1}{\sqrt n}\Bigg[\ket 0-e^{i\pi/4}\sum_{j=1}^{d-1}\ket j\Bigg]\,,\\
\begin{split}
\ket{\psi^{2k-1}}&=
\frac{1}{\sqrt n}\Bigg[
\ket 0+\sqrt{\frac{n}{2}}\ket k-\frac{e^{i\pi/4}}{\sqrt n+1}\sum_{j=1}^{d-1}\ket j
\Bigg]\\
&=\frac{1}{\sqrt n}\Bigg[
\ket 0-e^{i\pi/4}\bigg(z\ket k+\frac{1}{\sqrt n+1}\sum_{k\ne j=1}^{d-1}\ket j\bigg)
\Bigg]\,,
\qquad k=1,\ldots,d-1,
\end{split}\\
\begin{split}
\ket{\psi^{2k}}&=
\frac{1}{\sqrt n}\Bigg[
\ket 0+i\sqrt{\frac{n}{2}}\ket k-\frac{e^{i\pi/4}}{\sqrt n+1}\sum_{j=1}^{d-1}\ket j
\Bigg]\\
&=\frac{1}{\sqrt n}\Bigg[
\ket 0-e^{i\pi/4}\bigg(z^*\ket k+\frac{1}{\sqrt n+1}\sum_{k\ne j=1}^{d-1}\ket j\bigg)
\Bigg]\,,
\qquad k=1,\ldots,d-1,
\end{split}
\end{align}
\end{subequations}
where
\begin{align}
z=\frac{1}{\sqrt n+1}-\sqrt{\frac{n}{2}}\,e^{-i\pi/4}
=\frac{1}{\sqrt n+1}-\frac{\sqrt n}{2}+i\frac{\sqrt n}{2}\,.
\end{align}

For qubits ($d=2$), we have $n=3$ and $z=e^{i2\pi/3}$.  The resulting POVM vectors are
\begin{subequations}
\begin{align}
\ket{\psi^0}&=\frac{1}{\sqrt3}\big(\ket0-e^{i\pi/4}\ket1\big)\,,\\
\ket{\psi^1}&=\frac{1}{\sqrt3}\big(\ket0-e^{i\pi/4}e^{i2\pi/3}\ket1\big)\,,\\
\ket{\psi^2}&=\frac{1}{\sqrt3}\big(\ket0-e^{i\pi/4}e^{-i2\pi/3}\ket1\big)\,,
\end{align}
\end{subequations}
which make up a trine measurement with the three POVM vectors in the equatorial plane of the Bloch sphere.  For $d>2$, $z$ is not a phase, so the PFSIC works by adjusting both the amplitude and phase of one component of each POVM vector relative to all the other components.

The trine measurement cannot distinguish two pure states that are related by a sign flip of the $z$ component of the Bloch vector.  The trine measurement is locally informationally complete for pure states near the north pole of the Bloch sphere, but it is not globally informationally complete.  This exemplifies the sort of ambiguity that prevents Fisher-symmetric measurements from being globally informationally complete, even though they are locally informationally complete for pure states near the fiducial state.

\subsection{II.~Minimal-plus-one PFSIC consisting of two orthonormal bases}

Here we construct a PFSIC from two orthonormal bases.  The resulting PFSIC POVM has $2d$ outcomes, one more than minimal, and can be implemented by flipping a fair coin to choose between measurement in the two bases.

Our objective is to find $2d-1$ orthonormal vectors, $\{\vec b_0,\sqrt2\vec b_j,\sqrt 2\vec c_j\}_{j=1,\ldots,d-1}$ in $n=2d$ dimensions, with $\vec b_0$ having all positive components.  We begin with any $d$-dimensional (real) orthonormal basis $\{\vec u_j\}_{j=0,\ldots,d-1}$, where
\begin{align}
\vec u_j=\sum_{\xi=0}^{d-1}u_j^\xi\vec e_\xi\,,
\end{align}
with $\vec u_0$ chosen to have all positive components, i.e., $u_0^\xi>0$, $\xi=0,1,\ldots,d-1$.  That these vectors are orthonormal means that the components $u_j^\xi$ form an orthogonal matrix.  Now define the ($2d$)-dimensional vectors
\begin{subequations}
\begin{align}
\vec b_0&=\frac{1}{\sqrt2}
\begin{pmatrix}
\vec u_0\\[2pt]
\vec u_0
\end{pmatrix}
=\frac{1}{\sqrt2}
\sum_{\xi=0}^{d-1}u_0^\xi\big(\vec e_\xi+\vec e_{d+\xi}\big)\,,\\
\vec b_j&=\frac{1}{\sqrt2}
\begin{pmatrix}
\vec u_j\\[2pt]
\vec 0_d
\end{pmatrix}
=\frac{1}{\sqrt2}
\sum_{\xi=0}^{d-1}u_j^\xi\vec e_\xi\,,\quad j=1,\ldots,d-1,\\
\vec c_j&=\frac{1}{\sqrt2}
\begin{pmatrix}
\vec 0_d\\[2pt]
\vec u_j
\end{pmatrix}
=\frac{1}{\sqrt2}
\sum_{\xi=0}^{d-1}u_j^\xi\vec e_{d+\xi}\,,\quad j=1,\ldots,d-1,
\end{align}
\end{subequations}
where $\vec 0_d$ denotes the $d$-dimensional vector of zeroes.  It is clear that these vectors are orthogonal and normalized as required.

Plugging the components of these vectors into the expression for the POVM vectors,
\begin{align}\label{eq:POVMvectors2}
\ket{\psi^\xi}
=\sum_{j=0}^{d-1} a^\xi_j\ket j
=b_0^\xi\ket 0+\sum_{j=1}^{d-1}(b^\xi_j+ic^\xi_j)\ket j\,,
\end{align}
we get, for $\xi=0,1,\ldots,d-1$,
\begin{subequations}\label{eq:chitau}
\begin{align}
\ket{\psi^\xi}&=\frac{1}{\sqrt2}
\Bigg(u_0^\xi\ket 0+\sum_{j=1}^{d-1}u_j^\xi\ket j\Bigg)
=\frac{1}{\sqrt2}\ket{\chi^\xi}\,,\\
\ket{\psi^{d+\xi}}&=\frac{i}{\sqrt2}
\Bigg(\mathord{-}iu_0^\xi\ket 0+\sum_{j=1}^{d-1}u_j^\xi\ket j\Bigg)
=\frac{i}{\sqrt2}\ket{\tau^\xi}\,.
\end{align}
\end{subequations}
Here the set $\{\ket{\chi^\xi}\}_{\xi=0,1,\ldots,d-1}$ makes up a real, orthonormal basis with positive zero components, $\big\langle0\big|\big.\chi^\xi\big\rangle=u_0^\xi>0$, and the set  $\{\ket{\tau^\xi}\}_{\xi=0,1,\ldots,d-1}$ makes up a closely related orthonormal basis, identical to the first except that all the zero components, $\big\langle0\big|\big.\tau^\xi\big\rangle=-i\big\langle0\big|\big.\chi^\xi\big\rangle=-iu_0^\xi$, have been rephased to be pure imaginary.

For a qubit, if we choose $\vec u_0=(\vec e_0+\vec e_1)\sqrt2$ and $\vec u_1=(\vec e_0-\vec e_1)/\sqrt2$, we have $\ket{\chi^0}=(\ket0+\ket1)/\sqrt2$ and $\ket{\chi^1}=(\ket0-\ket1)/\sqrt2$, i.e., the eigenstates of the Pauli $x$ operator, and $\ket{\tau^0}=-i(\ket0+i\ket1)/\sqrt2$ and $\ket{\tau^1}=-i(\ket0-i\ket1)/\sqrt2$, i.e., the eigenstates of the Pauli $y$ operator.  The POVM corresponds to flipping a fair coin to decide whether to measure in the Pauli $x$ or $y$ basis.

It is now obvious how to construct the POVM vectors~(\ref{eq:chitau}) directly in Hilbert space, without retreating to the real vector space: start with any real orthonormal basis whose zero components are nonzero, rephase the zero components to be positive, and then construct the bases $\{\ket{\chi^\xi}\}$ and $\{\ket{\tau^\xi}\}$.  It is clear that the basis $\{\ket{\chi^\xi}\}$ is used to estimate the real parts, $x^{k0}$, of the amplitudes of $\ket{\Psi(\vec x)}$, and the basis $\{\ket{\tau^\xi}\}$ is used to estimate the imaginary parts, $x^{k1}$.

We can construct a measurement that weights the real and imaginary parts differently by defining $2d$ POVM vectors $\ket{\psi^\xi}=\sqrt{p_\chi}\ket{\chi^\xi}$ and $\ket{\psi^{d+\xi}}=\sqrt{p_\tau}\ket{\tau^\xi}$, with $p_\chi+p_\tau=1$.  For this POVM, the coin flip that chooses between measurement in the two basis has probabilities $p_\chi$ and $p_\tau$.  Now the real vectors are given by
\begin{subequations}
\begin{align}
\vec b_0&=
\begin{pmatrix}
\sqrt{p_\chi}\vec u_0\\[2pt]
\sqrt{p_\tau}\vec u_0
\end{pmatrix}
=\sum_{\xi=0}^{d-1}u_0^\xi\big(\sqrt{p_\chi}\vec e_\xi+\sqrt{p_\tau}\vec e_{d+\xi}\big)\,,\\
\vec b_j&=\sqrt{p_\chi}
\begin{pmatrix}
\vec u_j\\[2pt]
\vec 0_d
\end{pmatrix}
=\sqrt{p_\chi}
\sum_{\xi=0}^{d-1}u_j^\xi\vec e_\xi\,,\quad j=1,\ldots,d-1,\\
\vec c_j&=\sqrt{p_\tau}
\begin{pmatrix}
\vec 0_d\\[2pt]
\vec u_j
\end{pmatrix}
=\sqrt{p_\tau}
\sum_{\xi=0}^{d-1}u_j^\xi\vec e_{d+\xi}\,,\quad j=1,\ldots,d-1,
\end{align}
\end{subequations}
and the classical Fisher matrix is diagonal,
\begin{subequations}
\begin{align}
\label{eq:00}
C_{j0,k0}&=4\vec b_j\cdot\vec b_k=4p_\chi\delta_{jk},\\
\label{eq:11}
C_{j1,k1}&=4\vec c_j\cdot\vec c_k=4p_\tau\delta_{jk},\\
C_{j0,k1}=C_{k1,j0}&=4\vec b_j\cdot\vec c_k=0.
\end{align}
\end{subequations}
The matrix elements~(\ref{eq:00}) give the Fisher information for measurements of the real parts, $x^{k0}$, of the amplitudes of $\ket{\Psi(\vec x)}$, and the matrix elements~(\ref{eq:11}) give the Fisher information for measurements of the imaginary parts, $x^{k1}$.  When $p_\chi=1$, the real parts can be estimated with the quantum-limited resolution that applies to measurements of each parameter separately, and the measurements provide no information about the imaginary parts.  When $p_\tau=1$, the roles of the real and imaginary parts are reversed.  When $p_\chi=p_\tau=\frac12$, we have a Fisher-symmetric measurement, for which the real and imaginary parts of the amplitudes are all measured with the same resolution, that being half the quantum-limited resolution with which each parameter can be measured separately.  Other values of $p_\chi=1-p_\tau$ give other tradeoffs, within the GM~bound, between measuring the real and and imaginary parts.  This discussion thus illustrates the sense in which the GM~bound can be thought of as an uncertainty principle for measuring the parameters that specify a pure state.
\end{widetext}


\begin{thebibliography}{10}
\bibitem{mixed}
In subsequent papers, we develop the theory of Fisher-symmetric informationally complete measurements for mixed states.
\end{thebibliography}
\end{document}